# Field-tunable partial antiferromagnetism, glassy spin dynamics, and magnetodielectric coupling in the quasi-one-dimensional spin-chain compound $Ca_3CoIrO_6$

Priyanka Mahalle[1,2], A. Kumar[1,2,*], Kumar Bharti[3], Dipanshu Bansal[3], P. D. Babu[4], and S. M. Yusuf[1,2,†]

[1]*Solid State Physics Division, Bhabha Atomic Research Centre, Mumbai-400085, India*
[2]*Homi Bhabha National Institute, Anushaktinagar, Mumbai-400094, India*
[3]*Department of Mechanical Engineering, Indian Institute of Technology Bombay, Mumbai 400076, India*
[4]*UGC-DAE Consortium for Scientific Research, Mumbai Centre, BARC Campus, Mumbai 400 085, India*

Corresponding authors: *amitkr@barc.gov.in, †smyusuf@barc.gov.in

**Abstract:**

We report a comprehensive investigation of the quasi-one-dimensional spin-chain compound $Ca_3CoIrO_6$ (CCIO) using a combination of structural, magnetic, thermodynamic, transport, Raman, and dielectric measurements. Temperature-dependent neutron powder diffraction confirms the rhombohedral R-3c structure down to 5 K without any structural phase transition. DC magnetization, ac susceptibility, and relaxation measurements reveal a gradual evolution from a high-temperature paramagnetic-like state to a partially disordered antiferromagnetic (PDA) state below ~100 K, accompanied by slow cluster-like spin dynamics followed by a freezing transition near 30 K. Isothermal magnetic hysteresis $M(H)$ loops demonstrate partial chain freezing, while robust exchange bias is observed in field-cooled protocols, highlighting the interplay between PDA ordering and frozen spins. Resistivity and specific heat data indicate strong coupling between spin and charge degrees of freedom, accompanied by activated transport behavior. Raman spectroscopy identifies pronounced anomalies in phonon frequencies and linewidths across multiple magnetic regimes, reflecting strong spin-lattice coupling. Polarization–electric field ($P$–$E$) measurements reveal temperature-dependent crossovers from linear dielectric to weakly hysteretic behavior, consistent with short-range polar correlations driven by spin-lattice interactions. These findings establish CCIO as a prototypical quasi-one-dimensional frustrated spin-chain system where geometric frustration, spin–orbit coupling, and low-dimensionality generate field-tunable PDA order, glassy spin dynamics, exchange bias, and magnetodielectric coupling. These results provide new insights into frustration-driven phases in low-dimensional oxides and point towards potential multifunctional applications based on intrinsic magnetodielectric and exchange bias phenomena.

## I.  Introduction:

The interplay of reduced dimensionality, magnetic frustration, and strong spin–orbit coupling (SOC) in transition metal oxides continues to yield unconventional ground states of broad interest in condensed matter research [1-6]. A particularly rich platform is provided by the $A_3BB'O_6$ family ($A$ = Ca, Sr; $B/B'$ = Co, Ni, Rh, Ir, Mn, Fe) which hosts quasi-one-dimensional chains of alternating $BO_6$ trigonal prisms and $B'O_6$ octahedra arranged on a triangular lattice [7-44]. The triangular arrangement of chains promotes competing interchain interactions, while the chain geometry enforces strong intrachain correlations. Antiferromagnetic (AFM) coupling between chains of Ising spins on a triangular lattice cannot be simultaneously satisfied, leading to a highly degenerate manifold of frustrated states. A characteristic outcome of this competition is the partially disordered antiferromagnetic (PDA) state, first established in $Ca_3Co_2O_6$ [7-13,43,45-48]. In this phase, two-thirds of the magnetic chains order antiferromagnetically while the remaining one-third remain disordered, reflecting the compromise between ferromagnetic (FM) intrachain and AFM interchain interactions. The PDA state is accompanied by stepwise magnetization plateaus and slow spin dynamics, both hallmarks of frustration in this family. The relative strengths of interchain and intrachain magnetic couplings can be fine-tuned by chemical substitution, particularly with 4d and 5d ions, where enhanced SOC modifies exchange pathways and anisotropy. In iridates, this synergy fosters spin-orbit entangled states, notably the $J_{eff} = 1/2$ Mott phase [49-51], which has broadened interest in this structural family as candidates for exotic correlated electrons phases [52-54].

In this context, $Ca_3CoIrO_6$ (CCIO) is of special interest. In this compound, $Co^{2+}$ ions occupy the trigonal prism sites and $Ir^{4+}$ ions occupy the octahedral sites, where the strong SOC of $Ir^{4+}$ ions strongly influences the exchange interactions and magnetic anisotropy. Previous studies of CCIO have reported spin-freezing behavior near $T_1 \sim 30$ K, but the existence and nature of PDA order have remained elusive [39-42]. In particular, discrepancies between magnetization, µSR, and density functional theory (DFT) studies have left open questions regarding whether CCIO realizes a PDA state, a conventional spin glass, or a short-range ordered phase [41,42].

In this work, we present a comprehensive experimental investigation of CCIO aimed at clarifying these important issues. Using dc and ac magnetometry, magnetization relaxation, specific heat, electrical transport, temperature-dependent Raman spectroscopy, and ferroelectric measurements, we uncover clear evidence for a PDA phase below a characteristic temperature $T_2 \approx 100$ K. Unlike the abrupt PDA transitions observed in related compounds, CCIO exhibits a gradual evolution from the paramagnetic-like to a PDA state, indicating that SOC and subtle structural effects act collectively to stabilize a progressive ordering process. Below $T_1 \sim 30$ K, the PDA phase freezes into a glassy state with persistent short-range magnetic correlations, giving rise to slow dynamics and unusual FM-like signatures.

Our results reconcile earlier experimental ambiguities by demonstrating that CCIO hosts both a PDA phase and glassy freezing, rather than a conventional spin glass alone. The findings highlight the key role of SOC in shaping the balance between order and disorder in geometrically frustrated spin chains. By establishing CCIO as a reference system where frustration, dimensionality, and relativistic effects converge, this study advances the understanding of partially disordered magnetism and opens pathways for engineering new correlated electron states in low-dimensional magnetic oxides.

## II. Experimental Details:

Polycrystalline CCIO was synthesized by the conventional solid-state reaction method. High-purity $CaCO_3$ (≥99.99%), $Co_3O_4$ (≥99.99%), and $IrO_2$ (≥99.99%) were mixed in stoichiometric proportions, calcined at 1273 K for 12 h in air, reground, pelletized, and subsequently annealed at 1373 K for 12 h. The final sintering was carried out at 1473 K for 60 h with intermittent grinding to ensure homogeneity and phase formation. All heating and cooling steps were performed at a rate of 3 K/min to prevent thermal stress and promote uniform phase formation. The phase purity and crystal structure were confirmed by X-ray diffraction (XRD). The Rietveld refinement was carried out using the FullProf Suite [55]. Neutron powder diffraction (NPD) measurements were performed on the PD-I diffractometer ($\lambda = 1.094$ Å) at the Dhruva Reactor, Bhabha Atomic Research Centre, Mumbai [56].

Magnetic measurements were carried out using a commercial vibrating sample magnetometer (VSM). Zero-field-cooled (ZFC), field-cooled cooling (FCC), and field-cooled warming (FCW) magnetization $M(T)$ curves were recorded between 5 to 300 K at various applied fields with a temperature sweep rate of 2 K/min. Isothermal magnetization loops $M(H)$ were measured under both ZFC and FCC conditions. After each measurement, the sample was warmed to 300 K and cooled to the desired temperature to eliminate magnetic history effects. Time-dependent dc magnetization (magnetic relaxation) measurements were performed using the VSM probe of a PPMS under FCC conditions at selected temperatures, following cooling at 2 K/min in a 1 kOe field, with field removal either immediately (60 s waiting time) or after 1800 s. In addition, ac susceptibility as a function of temperature was measured in the frequency range of 97–2497 Hz to investigate the dynamic magnetic response.

The electrical resistivity $\rho(T)$ was measured as a function of temperature using a two-probe setup integrated into a closed-cycle helium refrigerator. A polished CCIO pellet, with a thickness of 0.8 mm, was contacted with silver paste electrodes on opposite faces. Polarization versus electric field ($P$-$E$) loops were obtained using a ferroelectric analyser at selected temperatures. The specific heat measurements were carried out between 1.8 and 305 K in different magnetic fields using a Quantum Design Physical Property Measurement System with heat-capacity option.

Raman spectroscopy was performed in backscattering geometry using a Renishaw inVia confocal spectrometer with a 532 nm laser excitation (power < 13 mW) and a 50X objective. The spectral resolution was better than 1 $cm^{-1}$ using 2400 grooves/mm grating. Temperature-dependent measurements between 15 and 300 K were carried out using a He-exchange cryostat with vacuum better than ~$10^{-6}$ mbar and temperature stability of ± 0.1 K. The spectrometer was calibrated against the 521 $cm^{-1}$ mode of Si. Spectra were taken in the range 100 to 1250 $cm^{-1}$.

## III. Results and Discussion

### 1. Crystal Structure

The crystal structure of CCIO was investigated by XRD and NPD, as shown in Figs. 1(a)-(b). Rietveld refinement of the room-temperature diffraction patterns using the FullProf Suite [55] confirms that CCIO adopts the rhombohedral $K_4CdCl_6$-type crystal structure (space group *R*–

3c). The refined lattice parameters ($a = b = 9.209(2)$ Å, $c = 10.892(3)$ Å, and $V = 800.18(4)$ Å³) from XRD are consistent with previous reports [39,41]. NPD refinement is fully consistent with the XRD results, demonstrating the reliability of the structural determination.

From NPD refinements, Ca, Co, Ir, and O occupy the 18$e$, 6$a$, 6$b$, and 36$f$ Wyckoff positions, respectively, with six formula units per unit cell ($Z = 6$). Bond valence sum (BVS) analysis [57,58], a method to determine the oxidation state of a central metal atom, supports the presence of $Co^{2+}$ and $Ir^{4+}$ oxidation states, consistent with the earlier x-ray photoemission results [33]. These oxidation states correspond to $Co^{2+}$ ($3d^7$) and $Ir^{4+}$ ($5d^5$) electronic configurations, which strongly influence the magnetic and electronic properties of CCIO.

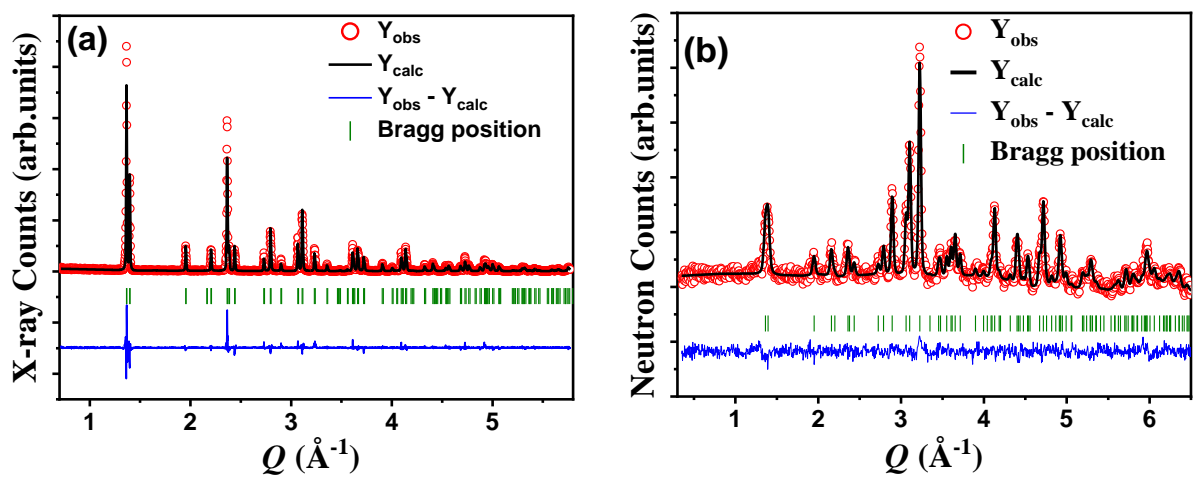

**Figure 1**: Rietveld refined (a) x-ray and (b) neutron diffraction patterns of $Ca_3CoIrO_6$ at 300 K.

The crystal structure consists of one-dimensional chains of alternating $CoO_6$ trigonal prisms and $IrO_6$ octahedra aligned along the $c$-axis. These chains are linear with nearly 180° Co-O-Ir bond angles, and Co–Ir separations of 2.720(3) Å, forming strong intrachain exchange pathways. The chains are arranged on a triangular lattice in the $ab$-plane, separated by Ca ions, with adjacent chains rotated by 120°, giving rise to a magnetic lattice that can support geometrical frustration (Fig. 2).

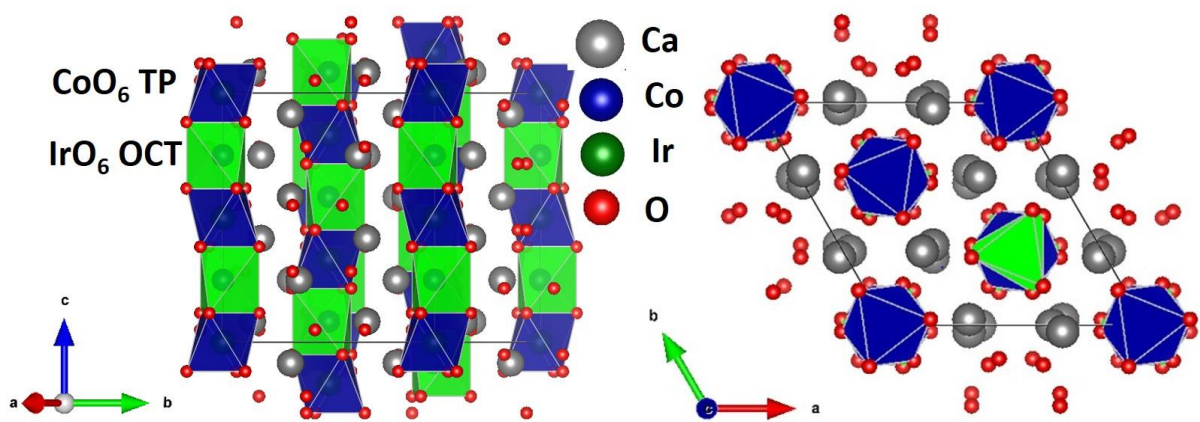

**Figure 2:** Crystal structure of $Ca_3CoIrO_6$, showing the alternating face-sharing $CoO_6$ trigonal prisms (TP) and $IrO_6$ octahedra (OCT) along the $c$-axis in the $R$-3$c$ space group. The triangular arrangement of chains in the $ab$ plane is also displayed.

Temperature dependent NPD patterns collected over 300 to 5 K [Fig. 3(a)] show no additional Bragg peaks or significant changes in intensities, indicating the absence of structural phase transitions. No additional magnetic scattering, expected from a long-range antiferromagnetic order, was observed. Lattice parameters exhibit anisotropic thermal behavior [Fig. 3(b)]: the $a$-axis shows the conventional contraction on cooling, while the $c$-axis shows a subtle anomaly below 100 K. This anisotropy reflects the distinct lattice dynamics along the spin-chain ($c$) direction.

The structural topology establishes strong intrachain superexchange pathways via nearly linear Co-O-Ir bonds, while weaker interchain couplings, frustrated by the triangular arrangement, generate competing interaction. This combination underpins the quasi-one-dimensional and frustrated magnetic behavior characteristic of ground states of the $A_3BB'O_6$ family.

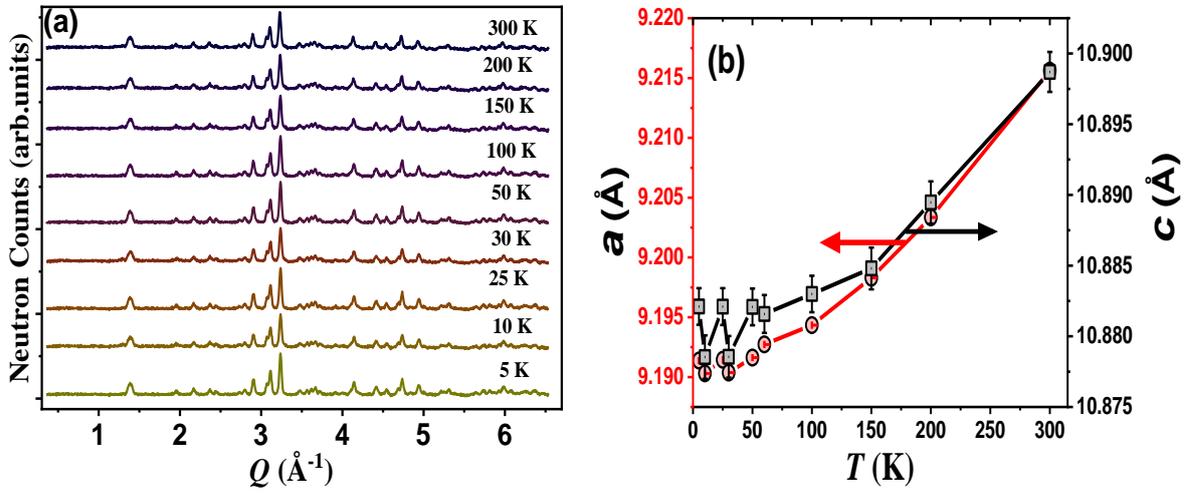

**Figure 3:** (a) Neutron powder diffraction patterns of $Ca_3CoIrO_6$ collected at representative temperatures. (b) Temperature dependence of the $a$ and $c$ lattice parameters.

## 2. DC Magnetization Study

### (i) Temperature-Dependent dc Magnetization ($M$ vs. $T$)

The temperature-dependent dc magnetization $M(T)$ curves, measured under ZFC and FCW protocols at various applied fields, are shown in Fig. 4(a). A clear bifurcation between FCC and ZFC curves appears below $T_1 \approx 30$ K, characteristic of a frozen spin-glass-like state. The transition temperature remains nearly field independent, and the persistence of bifurcation even at 50 kOe highlights the robustness of the frozen state. Under ZFC conditions, $M(T)$ drops sharply below 30 K, levels off between 20–30 K, and then rises slightly at lower $T$. In contrast, FCW curves show a broad hump centered near 27 K followed by a smooth increase down to 5 K, mirroring previous reports and confirming the intrinsic nature of the behavior [39-41,59]. Notably, at high fields (≥30 kOe), a broad and previously unreported anomaly develops in the temperature range of 55–90 K. This feature is reproducible in all measurements and appears exclusively in warming (ZFC and FCW) cycle, being absent in the cooling (FCC) data [Fig. 4 (b)]. Its presence underlines subtle and intrinsic field-induced correlations in the present system.

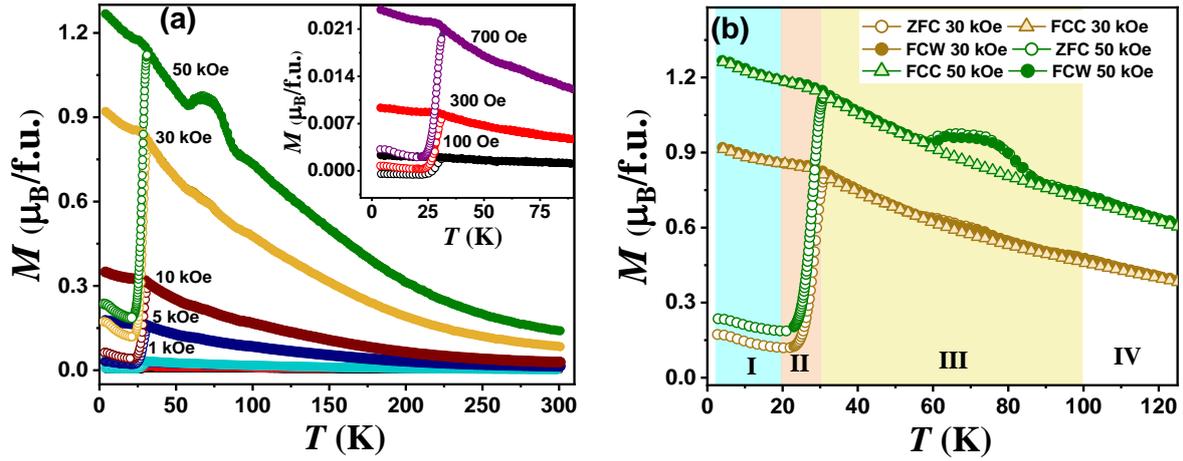

**Figure 4**: (a) Temperature dependence of magnetization $M(T)$ measured under ZFC and FCW protocols at various applied fields. Solid symbols represent the FCW while open symbols represent the ZFC data. Inset: enlarged view at low fields. (b) $M(T)$ curves at 30 and 50 kOe, highlighting the four distinct temperature regions discussed in the text.

The dc inverse susceptibility, $\chi^{-1}(T)$, measured at 1 kOe (ZFC), exhibits a linear temperature dependence consistent with the Curie-Weiss behavior above 250 K [Fig 5(a)], yielding $\mu_{eff}$ = 4.54 $\mu_B$/f.u. and $\theta_{CW}$ = 134.4(1) K. The positive $\theta_{CW}$ suggests dominant FM correlations, though DFT calculations [42] indicate intra- and inter-chain AFM couplings stabilized by SOC. The theoretical spin-only $\mu_{eff}$ [60] assuming $Co^{2+}$ ($S = 3/2$) and $Ir^{4+}$ ($S=1/2$) is 4.24 $\mu_B$/f.u., which is lower than the experimental value. The observed enhancement of 0.3 $\mu_B$/f.u. beyond the spin-only value is consistent with an unquenched orbital contributions arising from $Co^{2+}$ ($3d^7$, high-spin) and $Ir^{4+}$($5d^5$) ions. Moreover, FCC $\chi^{-1}(T)$ curves are field dependent [Fig. 5(b)], revealing short-range spin correlations persisting up to room temperature. Similar behavior was reported for $Sr_3CoIrO_6$ [24] due to SOC effects of $Co^{2+}$ $d^7$ states.

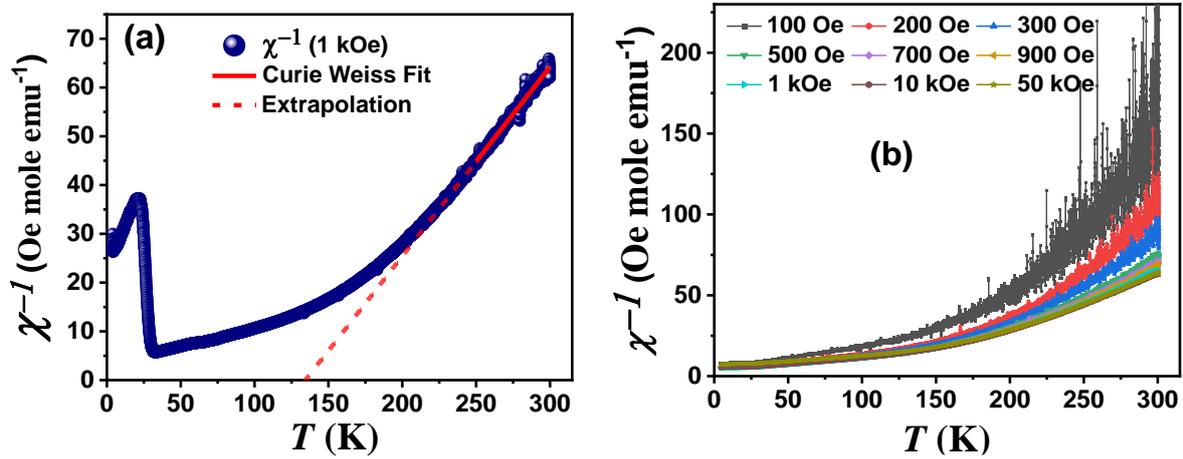

**Figure 5**: (a) Inverse dc susceptibility, $\chi^{-1}(T)$, measured at 1 kOe (ZFC), along with a Curie-Weiss fit (solid line). (b) $\chi^{-1}$ under the FCC protocol at different external fields.

The observed magnetic behavior can be divided into four distinct temperature regimes:

**Region I (5-20 K)**: The low-temperature spin-glass like state with frozen spins and increasing magnetization;

**Region II (20-30 K)**: The sharp spin-glass like freezing transition marked by bifurcation and peak features;

**Region III (30-100 K)**: Development of short-range correlations and magnetic fluctuations; broad hump at high fields (≥30 kOe) in the 55–90 K range; and

**Region IV (above 100 K)**: Field dependent dc susceptibility indicating short-range spin correlations.

More insight into such peculiar magnetic behavior has been obtained from isothermal magnetization study given below.

### (ii) Field-Dependent dc Magnetization ($M$ vs. $H$):

Distinct behaviors are observed in ZFC $M(H)$ plots across four temperature regimes, as shown in Fig. 6.

**Region I (5-20 K):** Between 2 and 20 K, $M(H)$ curves exhibit S-shaped loops with nearly linear high field behavior [Fig. 6(a)]. At 2-5 K, loops are symmetric and approach partial saturation. With increasing temperature (10 -15 K), the initial slope decreases and the response becomes more linear.

**Region II (20–30 K):** Rapid changes in $M(H)$ appear between 22 and 30 K [Fig. 6(b)]. Magnetization increases to ~1.32 $\mu_B$/f.u. at 70 kOe, compared to ~0.27 $\mu_B$/f.u. below 20 K. Pronounced loop opening is observed between 22 and 26 K, peaking at 24 K. At 24 K, magnetization was measured using both the scan mode, in which the magnetic moment is continuously recorded as the field is swept, and the step mode, in which the field is applied incrementally and allowed to stabilize at each value before measuring the magnetization [Fig. 6(c)]. These complementary measurement protocols confirm path-independent behavior. Above 27 K, loops regain the S-shape with narrowing width.

**Region III (30-100 K):** $M(H)$ loops show weak opening, negligible coercivity, and the moment at 70 kOe decreases steadily with temperature. The onset of nonlinearity in $M(H)$ curves is evident below 100 K [Fig. 6(d)]. There is a characteristic step-like feature near the origin of the $M(H)$ curves [Fig 6(d) inset]. As temperature decreases below 100 K, magnetization progressively develops a curvature at higher fields >30 kOe.

**Region IV (above 100 K):** At 200 K, the $M$-$H$ [Fig. 6(d)] loop exhibits an increasingly linear, paramagnetic-like response, though the system continues to deviate from ideal paramagnetic behavior due to the presence of short-range spin correlations. This interpretation is supported by inverse-susceptibility $\chi^{-1}(T)$ results, which reveal persistent short-range correlations up to room temperature.

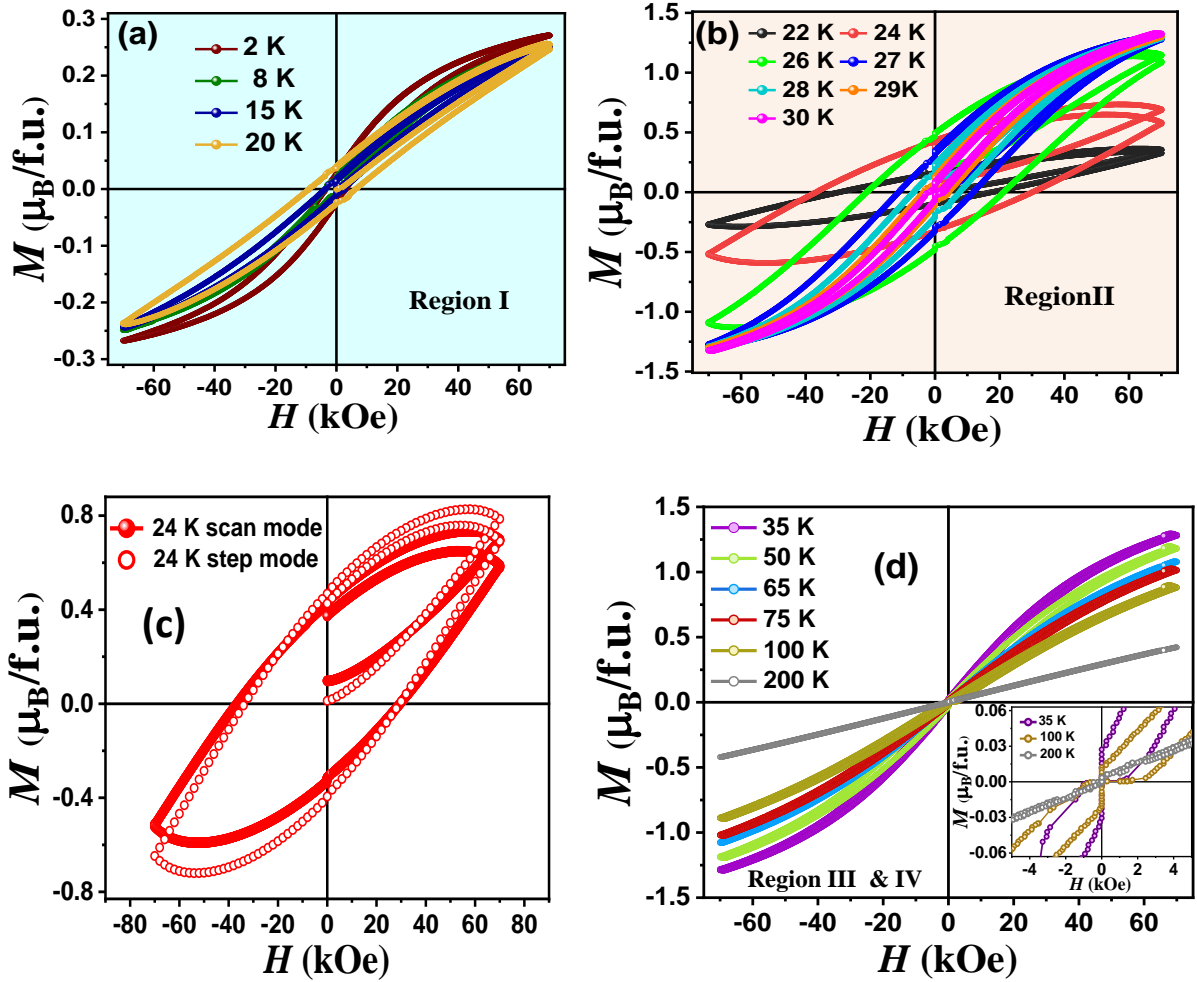

**Figure 6**: Isothermal $M(H)$ curves of CCIO measured under zero cooling field and up to ±70 kOe at (a) 2-20 K, (b) 22-30 K, (c) 24 K, and (d) 35-200 K. The inset of (d) highlights the step-like behavior near the origin for selected temperatures (35, 100 and 200 K).

The systematic evolution of the $M(T)$ (Fig. 4) and $M(H)$ (Fig. 6) curves across different temperature regimes reflects the intricate energy landscape of CCIO, where thermal fluctuations compete with magnetically frustrated interactions. This complex behavior can be rationalized within the framework of the partially disordered antiferromagnetic (PDA) model.

In the $A_3BB'O_6$ family [24], three magnetic models have been proposed: (i) incommensurate amplitude-modulated AFM order [7]; (ii) PDA order (two-thirds of chains ordered antiferromagnetically, one-third disordered) [14,15]; and (iii) ferrimagnetic order (two-thirds ferromagnetic, one-third antiferromagnetic) [61]. Among these, neutron diffraction and magnetization studies consistently identify the PDA configuration as the most relevant in $A_3BB'O_6$ compounds.

The magnetic behavior of such spin chain systems is characterized by two principal temperatures, $T_1$ and $T_2$. Below $T_2$, a cooperative PDA order establishes, accompanied by nonlinear magnetization behavior, while below $T_1$ ($< T_2$), a frozen or glassy magnetic state develops. In $Ca_3Co_2O_6$, $T_2$ occurs near 25 K [9,10] but increases significantly to about 90 K upon partial substitution of $Co^{3+}$ by $Rh^{4+}$ in $Ca_3CoRhO_6$ [14,16]. For $Sr_3CoIrO_6$ [24] and $Sr_3NiIrO_6$ [25,29], the PDA transition appears around 90 K and 85 K, respectively, marked by a sharp rise in magnetization below $T_2$. In contrast, CCIO exhibits a more gradual and field-dependent paramagnetic-to-PDA crossover, reflecting enhanced competition among exchange interactions, magnetic anisotropy, and spin–orbit coupling.

Lefrançois et al. [27] discussed an amplitude-modulated antiferromagnetic arrangement for $Sr_3NiIrO_6$, featuring completely compensated chains with spin dynamics governed by domain-wall nucleation, whereas Singleton et al. [26] argued that the PDA framework more accurately captures the magnetization behavior in $Sr_3NiIrO_6$ and $Sr_3CoIrO_6$. In analogy, the PDA model also provides the most consistent description for CCIO. The broad hump observed in $M(T)$ between 55–90 K (Fig. 4) can be attributed to a field-induced alignment of the magnetically disordered 1/3$^{rd}$ chain in the PDA picture.

The $M(H)$ curves of CCIO closely resemble those of $Sr_3CoIrO_6$ and $Sr_3NiIrO_6$ [26,30,62], reinforcing the PDA interpretation. The temperature-dependent evolution of the loops provides a direct evidence of a PDA-type ground state. Unlike the sharp metamagnetic steps typical of other $A_3BB'O_6$ compounds, CCIO exhibits a smooth, unsaturated magnetization with field, signifying enhanced frustration due to Co–Ir interactions. The onset of nonlinearity in the $M(H)$ curves below 100 K [Fig. 6(d)] and the progressive development of curvature at higher fields above 30 kOe signify the emergence of PDA correlations. The step-like hysteresis near zero field between 30–100 K [Fig. 6(d) inset] likely represents the reversible orientation of disordered chains characteristic of the PDA state. Similar field-dependent hysteresis behavior has also been observed in $Ca_3CoMnO_6$ within the field range of 20 to 90 kOe, indicating a spin-flip type transition in that material [63,64]. Between 20 and 30 K, large hysteresis with finite coercivity indicates slow spin dynamics and a transition to a frozen-PDA regime, consistent with observations in $Sr_3CoIrO_6$ [62], $Sr_3NiIrO_6$ [24,26,30], and $Ca_3CoRhO_6$ [16]. This behavior reflects a crossover from dynamic to frozen order, controlled by cooperative relaxation of disordered chains within a frustrated lattice. The similarities among these systems point to a shared origin rooted in triangular-chain topology and spin–orbit-entangled $Ir^{4+}$ states [26,29], while CCIO exhibits a more gradual and field-tunable evolution.

Zhang et al. attributed CCIO's unusual magnetism to the interplay between $Co^{2+}$ Ising anisotropy and $Ir^{4+}$ SOC. $Co^{2+}$ ions in trigonal prisms exhibit strong uniaxial anisotropy along the $c$-axis, slightly relaxed by weak Jahn–Teller distortions [42], while $Ir^{4+}$ ions in octahedra generate SOC-driven frustration through anisotropic exchange. This competition gives rise to a gradual, field-sensitive PDA state distinct from the abrupt transitions seen in other members of the series. The anomaly in the lattice parameter along the $c$-axis below 100 K (Fig. 3b) coincides with the onset of PDA correlations, evidencing strong spin–lattice coupling along the Co–Ir chains. Such coupling is characteristic of low-dimensional frustrated magnets where lattice degrees of freedom stabilize complex spin configurations.

Overall, magnetization measurements define four different magnetic regions: (i) Region I (5-20 K): Frozen-PDA state; (ii) Region II (20-30 K): PDA-to-frozen crossover; (iii) Region III (30-100 K): PDA state with rapidly evolving dynamics, as further confirmed by ac susceptibility, and broad field-dependent hump appearing (55-90 K) in ZFC/FCW curves; (iii) Region IV: short-range magnetic correlations persisting up to room temperature. These observations demonstrate that the interplay of geometric frustration, SOC, and spin–lattice coupling stabilizes a distinctive, field-tunable PDA ground state in CCIO, setting it apart from classical members of the $A_3BB'O_6$ family.

To further understand the characteristics of the PDA state and underlying interactions in CCIO, results of complementary studies have been discussed in the following sections.

### 3. Magnetization Dynamics Study
   (i) AC Susceptibility:

AC susceptibility of CCIO provides detailed insights into its magnetic dynamics and glassy behavior. Both the real ($\chi'$) and imaginary ($\chi''$) components of ac susceptibility, measured at frequencies ($\nu$) ranging from 97 to 2497 Hz under a fixed ac magnetic field of 10 Oe, are shown in Figs. 7(a) and 7(b). Clear anomalies appear at $T_p \sim 45$ K in $\chi''$ and 51 K in $\chi'$ (for $\nu = 97$ Hz), both shifting to higher temperatures with increasing frequency. The frequency-dependent shift of $T_p$, the reduction in peak height, and the finite $\chi''$ signal are hallmarks of glassy dynamics [65-67].

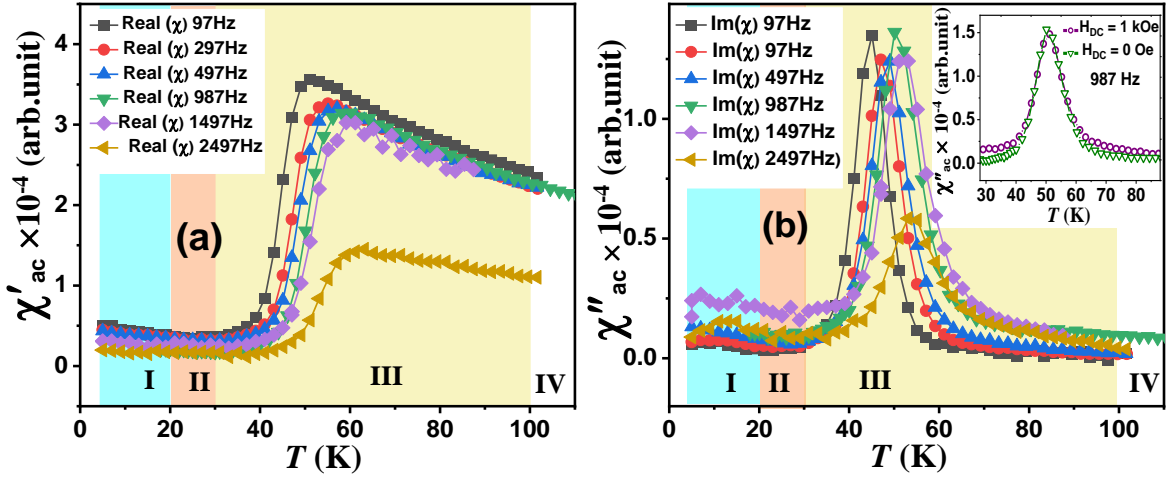

**Figure 7**: (a) Real ($\chi'$) and (b) imaginary ($\chi''$) parts of the ac susceptibility of CCIO measured at different frequencies as functions of temperature under an ac magnetic field of 10 Oe. Inset of (b) shows $\chi''$ measured with and without an applied dc field ($H_{DC} = 1$ kOe).

The frequency shift is quantified using the Mydosh parameter (S) [68,69],

$$S = \frac{\Delta T_f}{T_f \Delta \log_{10} f} \quad (1)$$

which yields $S = 0.12$ for CCIO. This value exceeds that of canonical spin glasses ($S \approx 0.005$–0.01) but falls within the cluster-glass regime ($S \approx 0.05$–0.18), while remaining below the superparamagnetic threshold ($S > 0.2$) [70]. Thus, the freezing involves spin clusters rather than isolated moments. The temperature dependence of the relaxation time ($\tau = 1/2\pi f$) was analyzed using Arrhenius, Vogel-Fulcher, and Power law forms [65,70,71]. Arrhenius law describes spin relaxation as thermal activation over energy barriers without interactions. At higher temperatures, spin relaxation follows the Arrhenius law, representing thermally activated, independent spin or cluster flipping over energy barriers, with relaxation time increasing exponentially as temperature decreases. The Arrhenius law:

$$\tau = \tau_0 \exp\left(\frac{E_a}{k_\beta T_f}\right) \quad (2)$$

yields $\tau_0 = 7.84 \times 10^{-12}$ s and $E_a = 0.073(3)$ eV, where, $\tau$ is the relaxation time, $\tau_0$ is the characteristic attempt time, $E_a$ is the activation energy, and $k_B$ is Boltzmann's constant. The plot of $\ln(\tau)$ vs $1/T_f$ shows a typical linear behavior which is evident from Fig. 8(a). The values of $\tau_0$ and $E_a$ are typical of thermally activated dynamics but neglecting inter-cluster interactions. As temperature decreases, interactions between spin clusters become non-negligible, leading to cooperative freezing dynamics that are better described by the Vogel-Fulcher law. The Vogel-Fulcher law is appropriate above or near the onset of collective freezing, characterizing the temperature regime where spins begin interacting and forming clusters. The Vogel-Fulcher law [72] is given by equation:

$$\tau = \tau_0 \exp\left(\frac{E_a}{k_B(T_f - T_0)}\right) \qquad (3)$$

where $T_0$ is the Vogel-Fulcher temperature, representing the temperature at which the relaxation time would diverge due to inter-cluster interactions. The fit, shown in Fig. 8(b), gives $\tau_0 = 2.52 \times 10^{-8}$ s, $E_a = 0.021(2)$ eV, and $T_0 = 22.7(5)$ K. The ratio $E_a/(k_B T_0) \approx 10.8$, which is much greater than 1, is a key indicator of cluster-glass behavior and is consistent with the $S$ parameter analysis. The condition $T_0 < T_{SG}$ further supports this assignment, while $T_0 \sim 2(E_a/k_B)$ suggests moderate inter-cluster coupling [73]. The Power law (critical slowing down) [72] describes approach to the spin-glass freezing, and is applicable closer to and below the spin-glass transition temperature. It is given by the equation:

$$\tau = \tau_0 \left(\frac{T_f - T_{SG}}{T_{SG}}\right)^{-zv} \qquad (4)$$

Here, $zv$ is the dynamic critical exponent, $\tau_0$ corresponds to the single spin-flipping time, $T_{SG}$ is the spin-glass transition temperature, and $v$ represents the critical exponent of the correlation length $\zeta = (T_f/T_g - 1)^v$. According to the dynamical scale hypothesis, $\tau$ relates to $\zeta$ as $\tau \sim \zeta^z$. The Power law fitting [Fig. 8(c)] yields $\tau_0 = 2.65 \times 10^{-6}$ s, $zv = 4.51(4)$, and $T_{SG} = 36.2(5)$ K. The relatively large $\tau_0$, far exceeding that of canonical spin glasses ($10^{-12}$–$10^{-13}$ s) [74], indicates slow and cooperative dynamics of spin clusters. The extracted $T_{SG}$ is consistent with the $M(T)$ data, which shows a spin glass transition near 30 K. These results, in line with previous Cole–Cole analysis [59], suggest dynamics intermediate between spin-glass and superparamagnetic systems, consistent with low-dimensionality and frustration.

The broadening distribution of relaxation times with decreasing temperature points to a system composed of superparamagnetic clusters or chain segments, rather than a conventional spin glass, which is also established from our analysis. The relaxation dynamics evolve from thermally activated independent spin flips (Arrhenius behavior, with $\tau_0 = 7.84 \times 10^{-12}$ s) at higher temperatures to collective cluster freezing captured by the Vogel-Fulcher law ($\tau_0 = 2.52 \times 10^{-8}$ s), and finally to critical slowing down near the spin-glass transition described by the Power law ($\tau_0 = 2.65 \times 10^{-6}$ s). This progression confirms that the system hosts interacting, frustrated spin clusters exhibiting slow, cooperative magnetic freezing consistent with a cluster-glass state.

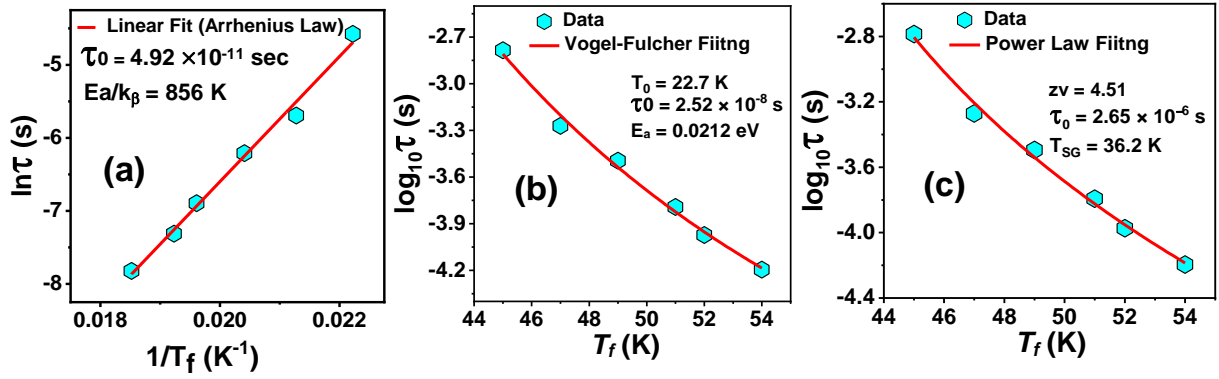

**Figure 8**: Temperature dependence of the relaxation time τ extracted from ac susceptibility measurements. The data are analyzed using (a) Arrhenius law, (b) Vogel-Fulcher law, and (c) power law (critical slowing down) models. Solid lines represent the corresponding fits.

Notably, the persistence of both χ′ and χ″ anomalies up to 1 kOe dc field, as shown in the inset of Fig. 7(b), unlike in $Ca_3Co_2O_6$ and $Ca_3CoRhO_6$ [16], further underscores the robust, frustration-driven nature of glassy state in CCIO. In fact, previous reports [39] indicate that the ac susceptibility peak of CCIO remains unaffected even under fields as high as 40 kOe. Interestingly, the anomaly in ac susceptibility (40–60 K) does not coincide with that of dc susceptibility (30 K). Similar behavior was observed in related compounds, such as $Ca_3Co_2O_6$ [1,46,75], $Ca_3CoRhO_6$ [16], $Ca_3Co_{2-x}Mn_xO_6$ [76], and $Sr_3NiIrO_6$, where the characteristic features in ac and dc susceptibilities occur at different temperatures. In these compounds, the dc susceptibility anomaly, observed as the bifurcation between ZFC and FCW $M(T)$ curves, marks the onset of thermomagnetic irreversibility associated with spin freezing temperature ($T_1$). In contrast the ac susceptibility peak reflects the evolving glassy dynamics of the PDA state. The $\chi''(T)$ peak occurs between $T_1$ and $T_2$ and disappears below $T_1$, suggesting that the spin dynamics originate from the transition at $T_2$ but slows down below $T_1$ beyond the detection limit of the ac susceptibility measurement timescale. In CCIO, $T_1$ and $T_2$ correspond to approximately 30 K and 100 K, respectively. The presence of well-defined χ′ and χ″ peaks across this temperature range provides further evidence for the PDA state in CCIO, consistent with the behavior reported for other members of the $A_3BB'O_6$ family.

In summary, the ac susceptibility of CCIO reveals field-insensitive, frequency-dependent glassy dynamics characteristic of a cluster-glass like state. The large Mydosh parameter, Vogel–Fulcher analysis, and power-law scaling all point to spin freezing governed by cooperative dynamics, arising from geometric frustration and competing AFM interactions in the spin-chain lattice.

### (ii) Magnetization Relaxation:

Magnetization relaxation provides key insights into the nonequilibrium dynamics of geometrically frustrated systems such as CCIO. To probe the dynamic magnetic behavior below 30 K, which lies beyond the frequency window of the ac susceptibility measurements, time-dependent magnetization measurements were performed at 5, 20, 23, 25, and 27 K, as shown in Fig. 9, covering Regions I and II of the phase diagram. For $T > 30$ K, relaxation could not be measured within this protocol because the relaxation timescales ($10^{-6}$–$10^{-12}$ s) fall beyond the experimental resolution and are instead captured in the ac susceptibility results discussed earlier.

The experimental protocol involved cooling at a rate of 2 K/min under a 1 kOe applied field (FCC condition), followed by field removal under two scenarios: (i) immediately after reaching the target temperature (60 sec waiting time) and (ii) after a waiting period of 1800 seconds. The subsequent decay of magnetization was recorded for 15,000 seconds. The relaxation curves were fitted to the stretched exponential model [77-79]:

$$M(t) = M_0 + M_g \exp\left(-\left(\frac{t}{\tau}\right)^\beta\right) \qquad (5)$$

Here, $M_0$ is the residual (non-relaxing) magnetization, $M_g$ the glassy component, $\tau$ the average relaxation time, and $\beta$ the stretching exponent ($\beta = 1$ corresponds to pure exponential decay, while $\beta < 1$ reflects a broad distribution of relaxation times typical of spin glasses). Representative fits are shown in Fig. 9(a), and the parameters are summarized in Table I.

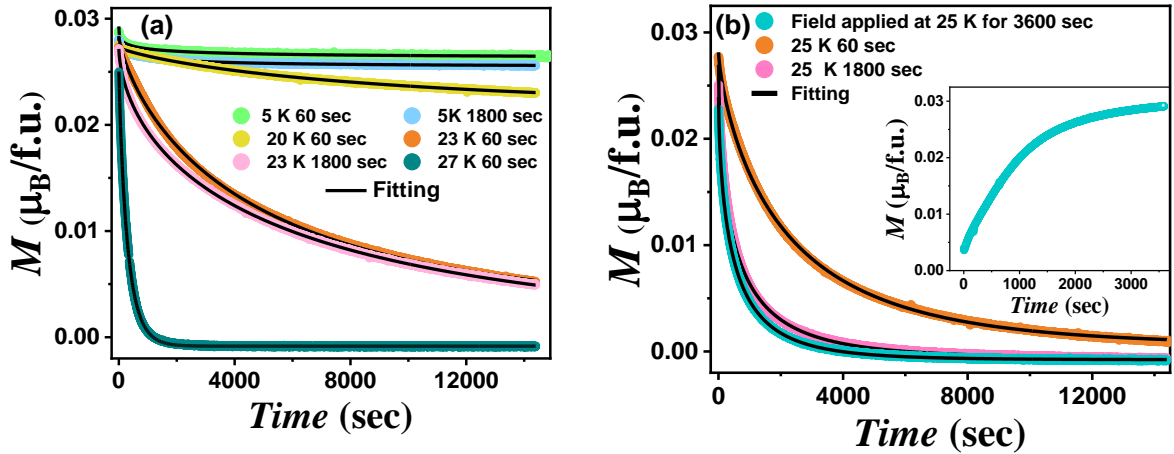

**Figure 9**: (a) Time evolution of magnetization ($M$) at 5, 20, 23, and 27 K, with data fitted to the stretched exponential model. (b) Magnetization relaxation at 25 K under an alternative protocol: the sample was cooled in zero field from 300 K to 25 K, a magnetic field was applied at 25 K for 1 hour (time evolution of $M$ shown in the inset), subsequently switched off to record the relaxation of $M$ as a function of time.

**TABLE I.** Fitting parameters obtained from stretched-exponential analysis of magnetization relaxation at different temperatures for two waiting times.

| $T$ (K) | Waiting Time (sec) | $M_0$ ($\mu_B$/f.u) | $M_g$ ($\mu_B$/f.u) | $\tau$ (sec) | $\beta$ |
|---|---|---|---|---|---|
| 5 | 60 | 0.02631 | 0.00294 | 493(2) | 0.326(5) |
| 5 | 1800 | 0.02554 | 0.00271 | 671(7) | 0.406(2) |
| 20 | 60 | 0.02159 | 0.00565 | 9656(4) | 0.819(2) |
| 23 | 60 | 0.00263 | 0.02577 | 4869(8) | 0.746(2) |
| 23 | 1800 | 0.0001 | 0.02724 | 5996(9) | 0.6065(5) |
| 27 | 60 | 0.0008 | 0.026 | 301.6(3) | 0.8999(1) |

The results reveal coexistence of frustrated magnetic order and glassy dynamics. At 5 K, $M_0$ is largest, indicating strong spin freezing, whereas $M_0$ decreases substantially, between 23-27 K, consistent with enhanced thermal fluctuations. Conversely, the glassy component $M_g$ grows with increasing temperatures ($\geq$ 23 K), suggesting that a larger fraction of the moment participates in relaxation as the system approaches the spin freezing temperature. The relaxation time $\tau$ shows aging: it increases with waiting time at low temperatures (5 K and 23 K), reflecting slow relaxation when the system is allowed to settle. Across all temperatures, the decay remains non-exponential ($\beta < 1$), with more stretched relaxation at low temperatures and faster but still glassy dynamics near the $T_1$. These behaviors highlight pronounced glassy characteristics below 30 K. In Region I ($T < 20$ K), the large $M_0$ ($\approx 0.026$ $\mu_B$/f.u, at 5 K) locks the system into a metastable, partially magnetized state. The increase in $\tau$ (493 → 671 s at 5 K) with waiting time further supports hierarchical energy landscapes typical of a spin-glass phase.

In Region II (20–30 K), where thermal fluctuations compete with freezing, relaxation becomes strongly temperature dependent. Here, small increases in temperature can unlock some frozen spins, leading to temperature-dependent changes in both the magnetization and the hysteresis loop. At 25 K, additional field-dependent relaxation measurements were conducted. The sample was first cooled to 25 K in zero field, then subjected to a magnetic field of 1 kOe. The system was allowed to evolve under this applied field for 3600 s, after which the field was removed, and magnetization relaxation was recorded as a function of time [Fig.9(b)]. The inset shows the time evolution of magnetization under the applied field, and fit parameters are summarized in Table II. Remarkably, relaxation curves converge to similar final states regardless of whether the field was applied during cooling or directly at 25 K, indicating that thermal history does not strongly affect the metastable configuration. These features mirror the memory erasure and hierarchical relaxation observed in structural glasses and disordered alloys, where minor perturbations disrupt metastable configurations.

**TABLE II.** Fitting parameters obtained from the stretched-exponential analysis of magnetization relaxation at 25 K under different measurement protocols.

| Measurement Condition | Waiting Time (sec) | $M_o$ ($\mu_B$/f.u) | $M_g$ ($\mu_B$/f.u) | $\tau$ (sec) | $\beta$ |
|---|---|---|---|---|---|
| Field applied at 300 K | 60 | 0.0008 | 0.02755 | 2301(2) | 0.7490(8) |
| Field applied at 300 K, sample cooled to 25 K and allow to dwell | 1800 | -0.0008 | 0.02824 | 552(1) | 0.5614(6) |
| Cooling in zero field from 300 K. Field applied at 25 K for 3600 s, and then removed | 60 | -0.0010 | 0.02515 | 505.4(6) | 0.6077(5) |

Overall, magnetization relaxation demonstrates that CCIO combines frozen and glassy magnetic components. Region I is dominated by strong spin freezing and aging effects, while Region II shows accelerated, history-independent relaxation near the $T_1$. These findings confirm that CCIO embodies the hallmark nonequilibrium dynamics of spin glasses, with metastability, aging, and memory erasure arising from the interplay of spin disorder, frustration, and spin-orbit coupling.

## 4. Exchange Bias:

Exchange bias is a phenomenon characterized by a shift of the magnetic hysteresis loop along the field axis, caused by interfacial exchange coupling that induces unidirectional anisotropy [80,81]. The EB field ($H_{EB}$) is quantified as:

$$H_{EB} = (H_+ + H_-)/2 \qquad (6)$$

where $H_+$ and $H_-$ are the positive and negative intercepts of the $M(H)$ loop on the field axis. By convention, when cooling in a positive field direction, a leftward (negative) shift of the hysteresis loop corresponds to $H_{EB} < 0$ [80]. EB is typically observed under the FCC conditions, but in CCIO it also emerges under the ZFC condition.

The $M(H)$ loop shift in CCIO is highly temperature sensitive. In the ZFC protocol, negative EB is observed between 5 and 28 K [Fig. 10(a)], while no EB is detected above 28 K. Instead, a step-like feature appears at the loop origin (> 30 K), as discussed earlier [Fig. 6(d)]. In addition to the horizontal loop shift, a vertical shift ($M_{EB}$) is observed, estimated as:

$$M_{EB} = (M_+ + M_-)/2 \qquad (7)$$

where $M_+$ and $M_-$ are the positive and negative intercepts of the loop on the magnetization axis. In the ZFC state, $M_{EB}$ exhibits an opposite trend to $H_{EB}$ [Fig. 10(b)]. At 24 K, both parameters reach maxima: $H_{EB} \approx 5.6$ kOe and $M_{EB} \approx 0.055$ $\mu_B$/f.u. The unusually large $H_{EB}$ at 24 K makes CCIO as a promising material for spintronic devices.

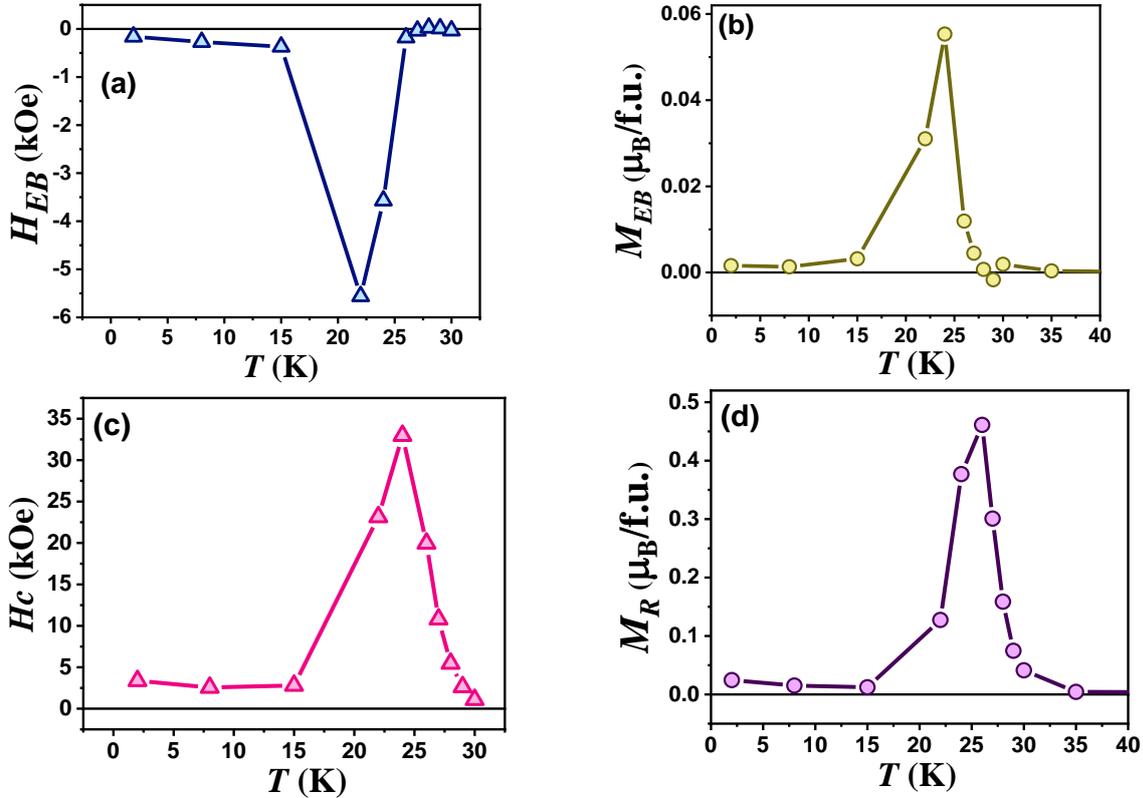

**Figure 10**: Temperature dependence of (a) vertical exchange bias ($H_{EB}$), (b) horizontal exchange bias ($M_{EB}$), (c) coercivity ($H_C$), and (d) remnant magnetization ($M_R$) for CCIO, measured under a maximum applied field of ±70 kOe following a zero-field-cooled protocol.

The coercivity ($H_C$) and remnant magnetization ($M_R$) are further extracted from the $M(H)$ loops using: $H_C = |H_+ − H_-|/2$ and $M_R = (M_+ − M_−)/2$. Their temperature dependences [Figs.10(c), (d)] are non-monotonic, peaking at 24 K ($H_C$) and 27 K ($M_R$), and vanishing above 30 K. Together, these features reveal giant EB behavior in Region II (20-30 K), coinciding with rapid dynamical changes in both $M(H)$ [Fig.6(b)] and $M(T)$ [Fig. 4].

Under FCC conditions (10 kOe), hysteresis loops exhibit vertical shifts at 2, 8, and 20 K in Region I [Fig. 11(a)]. At 8 K, increasing the cooling field from 10 kOe to 20 kOe enhances the vertical shift [Fig. 11(b)], consistent with complete spin freezing where it remains magnetized even after field removal. This behavior reflects the frozen PDA state and strong memory effects, corroborated by relaxation measurements [Fig. 9(a)], where a large remanent $M_0$ locks the system into a metastable, partially magnetized state Consequently, the vertical shift in the hysteresis loop under FC conditions in Region I is originates from this incomplete relaxation and spin alignment.

In contrast, the Regions II-III exhibit overlapping ZFC and FC loops with no vertical offsets [Fig.11(c)], indicating minimal memory effects beyond Region I. However, EB persists in Region II for both protocols, demonstrating its robustness against cooling history.

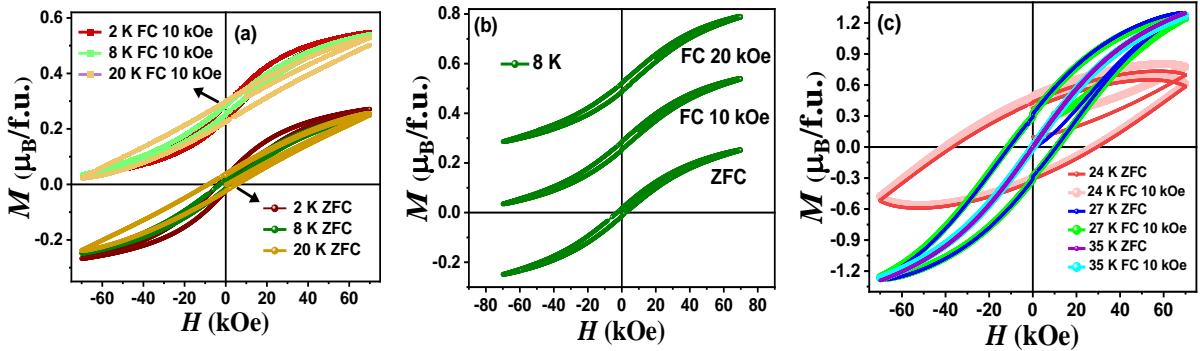

**Figure 11**: Isothermal magnetization (*M-H*) loops of CCIO measured under (a) zero-field-cooled (ZFC) and 10 kOe field-cooled (FC) conditions at 2, 8, and 20 K, (b) ZFC, 10 kOe, and 20kOe FC conditions at 8 K, and (c) ZFC and 10 kOe FC conditions at 24, 27, and 35 K.

The EB mechanism in CCIO arises from the interplay of spin-glass dynamics and AFM order within the PDA framework. The PDA structure in CCIO stabilizes the EB effect through the convolution of two energy landscapes: the highly degenerate spin-glass landscape and the AFM sublattice configuration. Below $T_1 \approx 30$ K, disordered Co–Ir chains undergo glassy freezing, while the remaining chains establish long-range AFM order. Interfacial coupling between these two regions pins uncompensated spins, producing unidirectional anisotropy and shifting the hysteresis loop. This represents a volume-based EB, stabilized by spin frustration rather than structural disorder. A similar PDA-stabilized EB was reported in $Sr_3NiIrO_6$ [30], and in $Sr_3CoIrO_6$ [24].

This cooperative phase coexistence resembles that in $Fe_xNbS_2$, where interplay of spin-glass and AFM orders leads to a giant EB [82]. Maniv *et al.* [82] attributed this to uncompensated spin-glass moments inducing a pronounced lateral shift of the hysteresis loop. In CCIO, Ising anisotropy from $Co^{2+}$ (trigonal prismatic) and strong spin-orbit coupling from $Ir^{4+}$ (octahedral) further enhance this effect. Unlike pyrochlore iridates, $A_2Ir_2O_7$ ($A$ = rare-earth, Y or Bi) [83], where EB originates from defect-driven FM droplets in an AFM matrix, CCIO retains robust

Ir$^{4+}$ valence as confirmed by our NPD study and BVS analysis. Thus, EB here is not defect-driven but emerges intrinsically from geometric frustration and spin–orbit entanglement.

Overall, CCIO exhibits an intrinsic and volumetric EB effect rooted in the coexistence of glassy and AFM phases in a geometrically frustrated quasi-1D magnet. Its unusually large magnitude, persistent under ZFC, and vertical shifts under FCC distinguish it from conventional FM/AFM heterostructures. These properties not only establish CCIO as a model system for volume-based EB but also highlights its promise for spintronic technologies, where stable and tunable EB is crucial.

### 5. Resistivity:

The electrical resistivity ($\rho$) of CCIO provides key insights into the charge transport mechanisms underlying its complex magnetic phase diagram. Resistivity measured between 35-100 K using two-probe method is shown in Fig. 12(a). CCIO exhibits typical insulating behavior, with $\rho$ increasing exponentially as $T$ decreases. The inset of Fig. 12 (a) shows $\rho$ on a logarithmic scale along with the temperature derivative (d$\rho$/d$T$). Notably, d$\rho$/d$T$ becomes negative below 60 K, reaches a minimum at 40 K and recovers towards 35 K. This anomaly around 40 K coincides with the ac susceptibility anomaly observed between 30 and 60 K (Fig. 7), where the imaginary part of ac susceptibility displays a frequency dependent peak between 40 and 60 K (Region III). These results suggest that the resistivity anomaly has a magnetic origin. The temperature dependence of $\rho$ can be described by an activation law:

$$\rho(T) = \rho_0 \exp\left(\frac{\Delta}{k_B T}\right) \quad (8)$$

where, $\rho_0$ is the pre-exponential factor, $\Delta$ the activation energy, $k_B$ the Boltzmann constant, and $T$ the absolute temperature. The *ln($\rho$)* vs $T^{-1}$ plot [Fig. 12 (b)] reveals three regions with distinct slops: $\Delta 1 \approx 0.06$ eV (100-80 K), $\Delta 2 \approx 0.02$ eV (75–50 K), and $\Delta 3 \approx 0.03$ eV (47–40 K). These values are comparable to those reported for $Ca_3Co_2O_6$ [84], reinforcing the role of specific electronic structure and magnetic interactions. Previous reports on CCIO show insulating behavior in the 100–400 K range, consistent with the isoelectronic nature of Co and Ir, leading to similar electronic characteristics as in $Ca_3Co_2O_6$ [85]. Remarkably, the activation energy $\Delta 2 \approx 0.02$ eV matches the value obtained from the Vogel-Fulcher analysis of ac susceptibility data in the same temperature range [Fig. 8(b)], suggesting a correlation between charge transport and spin dynamics. In summary, CCIO exhibits an insulating ground state with the resistivity anomalies closely tied to its magnetic transitions. The strong correspondence between charge transport and spin dynamics underscores the pivotal role of magnetism in governing the electronic properties of this material.

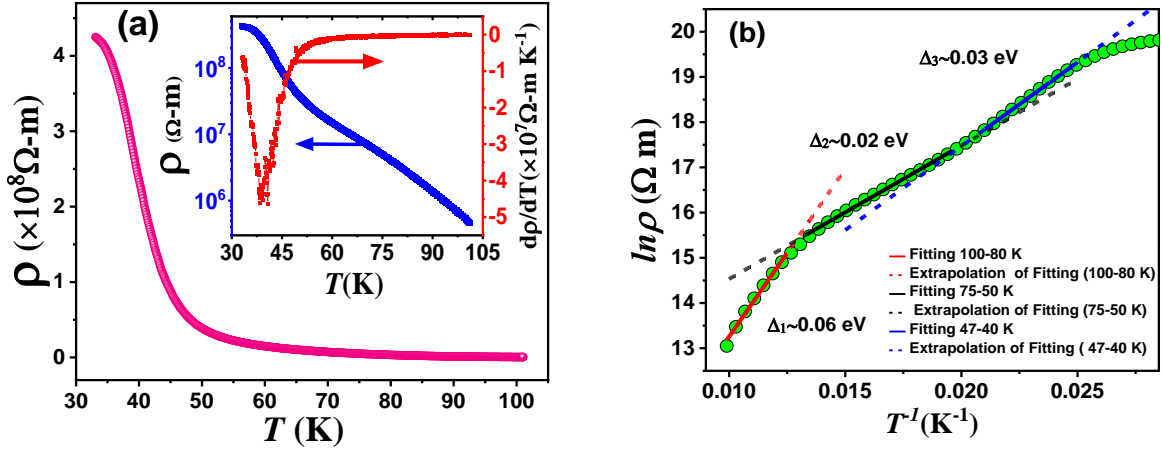

**Figure 12**: (a) Temperature dependence of resistivity (ρ) of CCIO measured on warming from 35 K to 100 K at a rate of 1 K/min. The inset shows ρ on a logarithmic scale together with the temperature derivative (d$\rho$/d$T$). (b) Plot of ln $\rho$ versus 1/$T$ in the range 35–100 K. The solid line is a fit to the activated transport law, while the dotted line represents the extrapolation of the fitted curve.

### 6. Specific Heat:

The specific heat at constant pressure ($C_p$) of CCIO at zero field as a function of temperature is shown in Fig. 13(a). At 300 K, the measured $C_p$ (~246.5 J/mol-K) remains below the classical Dulong–Petit limit of 274.36 J/mol-K (n = 11 atoms / f.u.), indicating incomplete phonon saturation at room temperature. To model the lattice contribution, data over 50 and 200 K were fitted using a Debye - Einstein heat capacity model [86] expressing the lattice specific heat $C_{latt}$ as:

$$C_{latt} = mC_p(\text{Debye}) + (1-m)C_p(\text{Einstein}) \qquad (9)$$

where m is the fractional contribution of the Debye model, and $C_p$(Debye) and $C_p$(Einstein) are the Debye and Einstein lattice specific heats, respectively, at constant pressure, given by:

$$C_{p(\text{Debye})}(T) = 9nR\left(\frac{T}{\Theta_D}\right)^3 \int_0^{\frac{\Theta_D}{T}} \frac{x^4 e^x}{(e^x - 1)^2} dx \qquad (10)$$

$$C_{p(\text{Einstein})}(T) = 3nR\left(\frac{\Theta_E}{T}\right)^2 \frac{e^{\frac{\Theta_E}{T}}}{(e^{\frac{\Theta_E}{T}} - 1)^2} \qquad (11)$$

Here, $\Theta_D$ and $\Theta_E$ represent the Debye and Einstein temperatures, respectively, accounting for acoustic and optical phonon contributions. However, the data could not be well fitted using only the lattice terms. Therefore, we introduced a linear temperature-dependent term:

$$C_p = C_{latt} + \gamma_{mag}T \qquad (12)$$

where the additional linear term reflects magnetic excitations rather than electronic contributions, consistent with the insulating nature of CCIO. The best fit yields $\Theta_D$ = 661(3) K, $\Theta_E$ = 214(1) K, $\gamma_{mag}$ = 52(1) mJ/mol-K², and m = 0.3, indicating dominant Debye contribution (~70%) with a significant Einstein component (~30%). The persistence of the linear term up to 300 K highlights robust quasi-one-dimensional magnetic correlations, in agreement with *H/M* analysis [Fig. 5(b)].

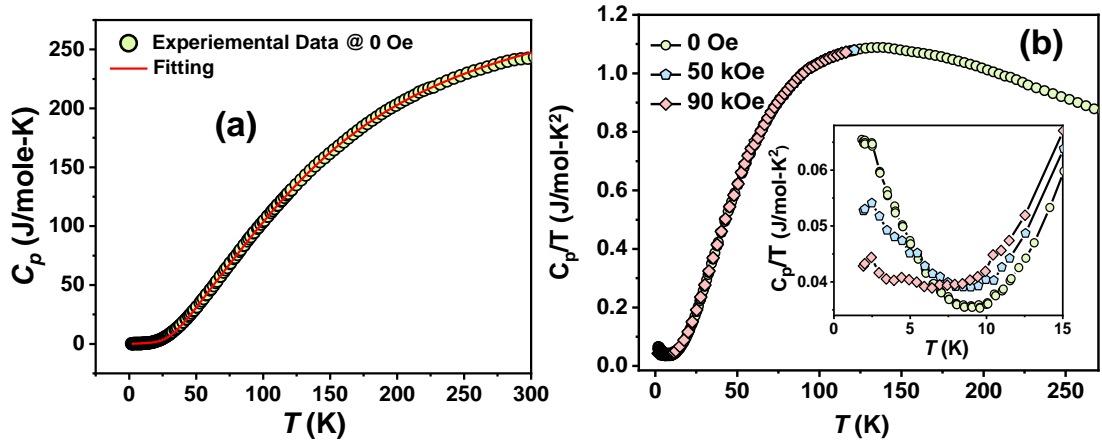

**Figure 13**: (a) Specific heat $C_p(T)$ of CCIO from 2 to 300 K. Solid line denotes the fitted lattice contribution using a combined Debye–Einstein model with an additional linear magnetic term from one-dimensional spin excitations. (b) $C_p/T$ vs $T$; inset shows an enlarged view of the low temperature region.

The plot of $C_p(T)/T$ vs. $T$ [Fig. 13(b)] shows a broad hump, consistent with the short range magnetic correlations reported in other $A_3BB'O_6$ oxides [17,75,87,88]. The inset of Fig. 13(b) shows an enlarged view of the low-temperature region, highlighting subtle features, such as low-lying spin excitations. A closer inspection below 10 K reveals a field-dependent upturn in $C_P(T)/T$, which develops more clearly at low temperatures. The amplitude of this feature decreases with increasing field, a trend consistent with Zeeman splitting of discrete energy levels often attributed to low-lying magnetic excitations or Schottky-like contributions, as observed in a spin liquid candidate honeycomb compound $H_3LiIr_2O_6$ [89]. After subtracting the lattice contribution, the derived magnetic specific heat, $C_M(T) = C_P(T) − C_{latt}(T)$ reveals a field independent anomaly around 30 K [Fig. 14 (a)], coinciding with the spin freezing transition. This anomaly likely arises from intrinsic low-energy excitations or spin cluster like dynamics associated with the magnetic glassy state. Additional features emerge between 50 and 65 K under applied fields [Fig. 14(b)], visible only during warming cycle, reflecting field induced reorientation of the partially disordered chains. This is consistent with features observed in the magnetization $M(T)$ data measured at 50 kOe [Fig. 4(b)].

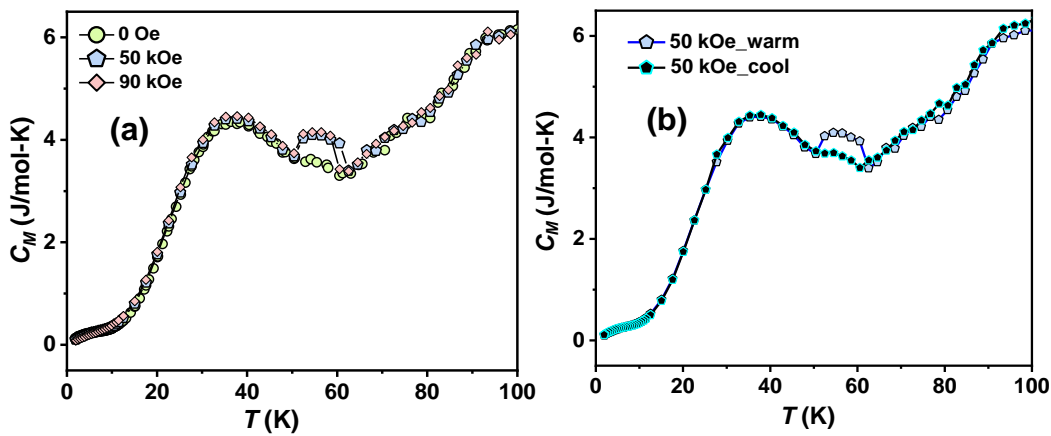

**Figure 14**: (a) Magnetic specific heat ($C_M$) as a function of temperature under zero, 50, and 90 kOe applied fields during the warming protocol. (b) Comparison of $C_M$ measured under 50 kOe during cooling and warming protocols.

Overall, the specific heat data demonstrate persistent magnetic excitations up to room temperature, a low-temperature Schottky anomaly linked to spin freezing, and field- and path-dependent features tied to frustrated spin-chain dynamics. These findings reinforce the quasi-one-dimensional and frustrated character of CCIO, consistent with related systems such as $Ca_3Co_2O_6$[75], $Ca_3CoRhO_6$ [17], and $Sr_3NiIrO_6$[24], where magnetic excitations persist well above the magnetic ordering temperature, as evidenced by inelastic neutron scattering measurements [27-29,90,91].

7. **Raman Spectroscopy:**

In the rhombohedral R-3c crystal structure, lattice vibrations follow well-defined symmetries that govern their Raman activity, thereby providing direct insight into atomic dynamics and bonding. Group theoretical analysis predicts 14 Raman active modes: 4 $A_{1g}$ and 10 $E_g$. The $A_{1g}$ modes involve symmetric vibrations of oxygen atoms with Co and Ir ions stationary, while doubly degenerate $E_{1g}$ modes correspond to in-plane vibrations primarily of Co ions within trigonal prisms, along with Ca and O displacements. One $E_g$ mode also features Ir as a nodal point. As shown in Fig. 15, the three dominant $A_{1g}$ modes mainly arise from the vibrations of the 12 oxygen atoms coordinated with Co and Ir.

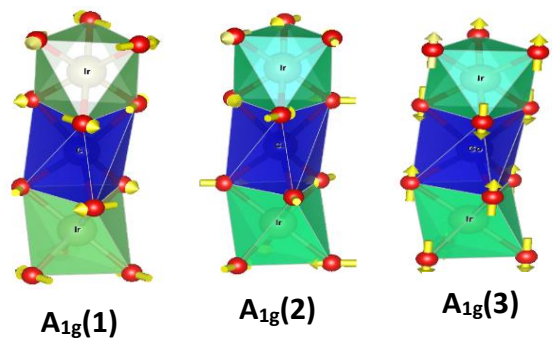

$A_{1g}(1)$    $A_{1g}(2)$    $A_{1g}(3)$

**Figure 15:** Visualization of three distinct $A_{1g}$ Raman-active vibrational modes in CCIO, primarily involving symmetric stretching motions of the 12 oxygen atoms coordinated around Co and Ir ions. Arrows indicate the directions of atomic displacements.

Fig. 16 (a) shows the raw Raman spectra of CCIO recorded at different temperatures, while baseline-corrected spectra for improved visualization of phonon modes is plotted in Fig. 16 (b). No additional peak appears down to the lowest measurement temperature (15 K), ruling out a crystallographic phase transition. With decreasing temperature, Raman peaks sharpen due to reduced anharmonic phonon-phonon interactions and suppressed thermal broadening. Notably, spectra at 200, 125, and below 24 K show enhanced intensity above 1000 cm$^{-1}$, suggesting magnetic scattering contributions. Fig. 16(c) represents the room temperature spectrum, highlighting four prominent peaks: A (400 cm$^{-1}$), B (450 cm$^{-1}$), C (680 cm$^{-1}$), and D (715 cm$^{-1}$). Lorentzian fits yield their frequencies and linewidths. Modes assignments are made by analogy with $Ca_3Co_2O_6$, which shares the same crystal symmetry and chain-like $CoO_6$ trigonal prism structure [92]. In $Ca_3Co_2O_6$, $A_{1g}$ modes (650–700 cm$^{-1}$) arise from symmetric oxygen stretching along the Co–O–Co chains, while the $E_g$ modes ($< 500$ cm$^{-1}$) involve Co–O bending and CoO$_6$ trigonal prism distortions [92]. Accordingly, peaks C and D in CCIO are attributed to $A_{1g}$ stretching modes, while peaks A and B correspond to $E_g$ bending/distortion modes involving Co$^{2+}$, Ca, and O displacements. Their slightly higher frequencies compared with Ca$_3$Co$_2$O$_6$ reflect stronger Ir$^{4+}$–O bonding due to Ir's smaller ionic radius and higher charge.

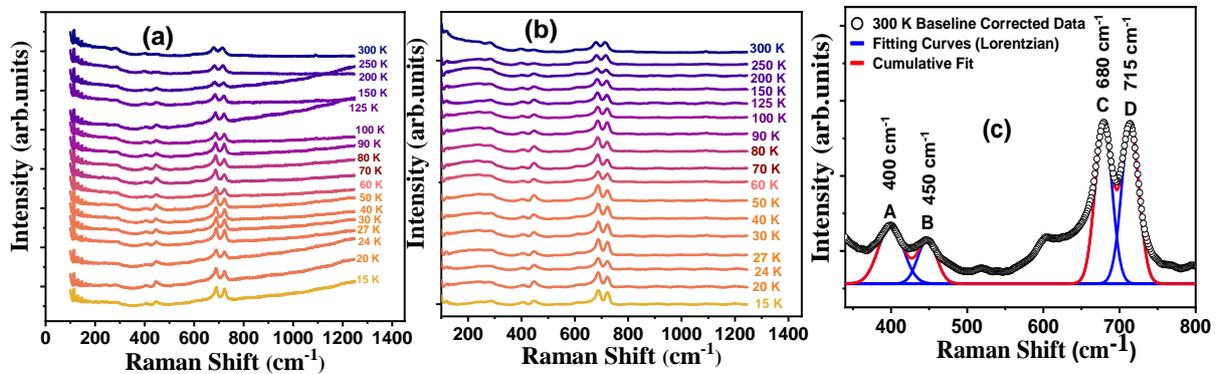

**Figure 16:** (a) Raw Raman spectra of CCIO collected at various temperatures, showing the evolution of phonon features with temperature. (b) Baseline-corrected spectra for improved visualization of temperature effects in peak positions and intensities. (c) Room temperature Raman spectrum with spectral deconvolution: the red line denotes the total fit using Lorentzian functions, while the blue lines indicate individual Lorentzian components corresponding to distinct Raman-active phonon modes.

Figures 17 (a-h) summarize temperature-dependent phonon frequencies ($\omega$) and linewidths ($\Gamma$) across four magnetic regions (I–IV). Distinct anomalies in both $\omega$ and $\Gamma$ closely correlate with CCIO's magnetic transitions, evidencing strong spin-lattice coupling. In Region IV (100–300 K), all modes generally harden with cooling, though peaks A and B deviate from this trend below 150 K, signaling onset of magnetic correlations. Simultaneously, linewidths ($\Gamma$) decrease monotonically across Region IV (Figs. 17(g–h)). In Region III (60–100 K), where µSR [41] detects static magnetic order, non-monotonic phonon trends emerge: peaks A and B harden and then soften, while peak C softens then hardens, reflecting a magnetostrictive strain. Linewidth anomalies further support enhanced spin–phonon coupling. Magnetization data [Fig. 4(b)] show FCC-FCW bifurcation and broad hump, reinforcing this interpretation. In Region III (30–60 K), where ac susceptibility shows frequency-dependent peaks, peaks A and C harden while their linewidths decrease, indicating slowing spin dynamics and short-range correlations. Region II (20–30 K), associated with spin-glass freezing, exhibits $\omega$ and $\Gamma$ anomalies for peaks A and B, consistent with glassy spin clusters distorting the lattice and impending coherent phonon propagation.

Dynamic Jahn-Teller (JT) distortions and SOC further enrich the phonon response. DFT calculations predict weak JT distortion at Co$^{2+}$ sites [42], insufficient for static long-range order but capable of inducing dynamic lattice fluctuations. The non-monotonic behavior of $E_g$ modes (A and B) below 150 K, including softening in Region III (60–100 K), aligns with JT-induced lattice strain. The absence of new Raman peaks rules out static JT distortion, instead supporting dynamic JT effects, moderated by strong Ir$^{4+}$ SOC, which enforces easy-axis anisotropy along the chains.

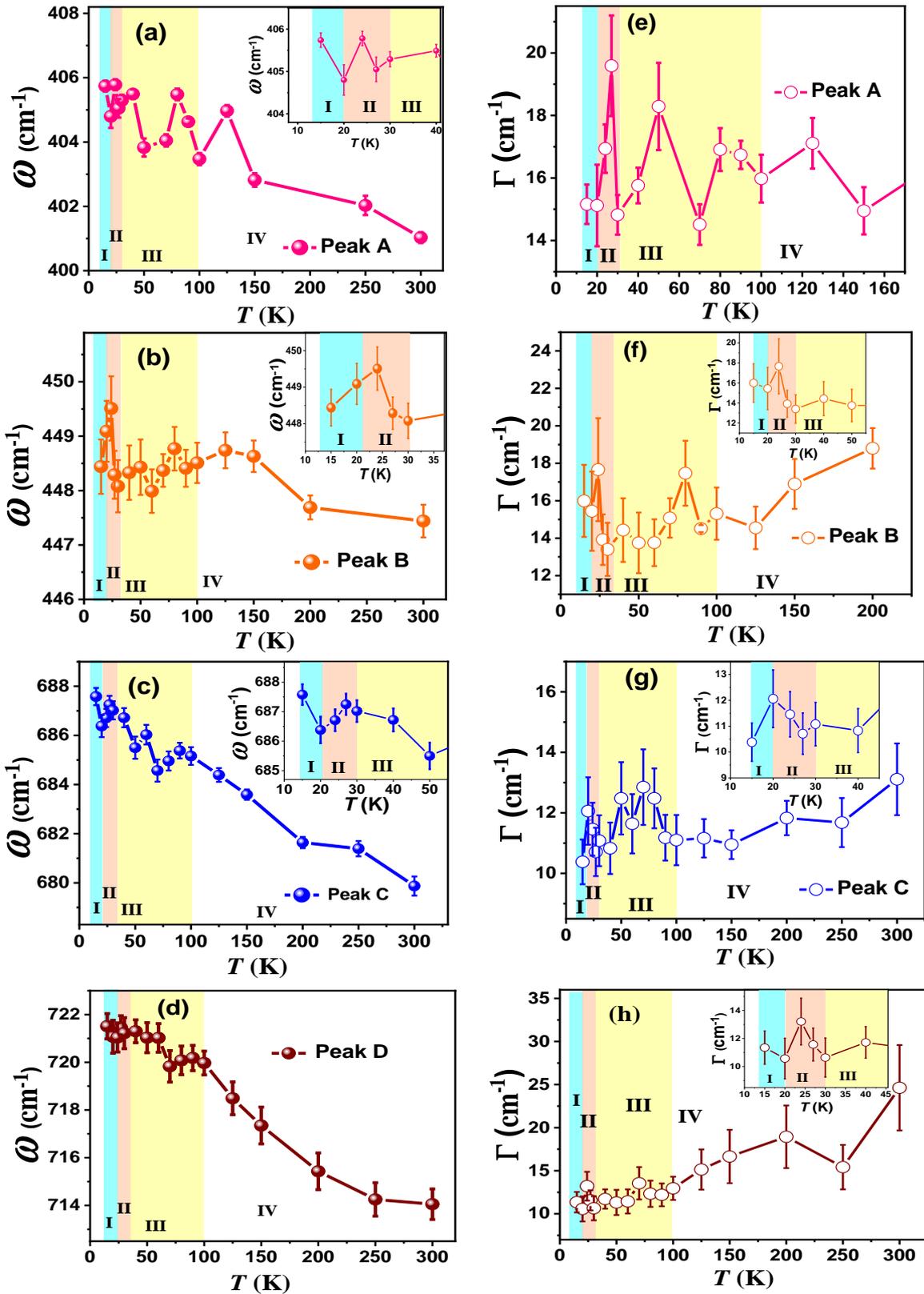

**Figure 17:** Temperature dependence of the phonon frequency ($\omega$) for (a) Peak A, (b) Peak B, (c) Peak C, and (d) Peak D, and phonon linewidth ($\Gamma$) for (e) Peak A, (f) Peak B, (g) Peak C, and (h) Peak D.

Enhanced Raman intensity above 1000 cm$^{-1}$, particularly at 200, 125, and below 24 K [Fig. 16(a)] points to magnetic scattering rather than electronic origins. Analogous broad high-frequency features in Ca$_3$Co$_2$O$_6$ (1200–3000 cm$^{-1}$) have been linked to spin disorder and frustration [92]. CCIO resistivity [Fig. 12(a)] confirms an insulating state, excluding charge density waves. This interpretation aligns with Gohil et al.'s conclusion for Ca$_3$Co$_2$O$_6$ [20], where such broad Raman features were attributed to the signatures of a disordered magnetic ground state, as supported by literature reported µSR measurements. In the case of CCIO, µSR measurements [41] reveal persistent dynamic spin disorder above 100 K, explaining the high-temperature intensity gain in the Raman spectra. Below ~30 K, additional enhancement reflects incomplete spin freezing and coexistence of glassy clusters with short-range AFM order. SOC driven spin-phonon coupling is essentially evident below 24 K (Fig.16(a)), where Ir$^{4+}$ anisotropy boosts high-frequency spectral weight.

In summary, Raman spectroscopy reveals that CCIO exhibits strong spin-phonon coupling, dynamic JT distortions (Co$^{2+}$), and strong SOC (Ir$^{4+}$) effects. Eg modes act as magnetic probes, displaying non-monotonic shifts and linewidth anomalies across transitions, while A$_{1g}$ modes remain structural fingerprints. The absence of structural transitions in NPD contrasts with clear Raman anomalies, showing that magnetic interactions dominate phonon behavior. Enhanced spectral weight above 1000 cm$^{-1}$ further highlights the role of spin disorder, SOC, and frustration in shaping CCIO's vibrational dynamics. These results position CCIO as a model 3d-5d mixed oxide, where lattice, spin and orbital interactions intertwine to produce complex and tunable quantum magnetic phases.

### 8. P–E Measurements

The dielectric behavior of CCIO exhibits a pronounced temperature dependence, as revealed by polarization-electric field (*P–E*) measurements [Figs. 18 (a-c)]. At low temperatures (15–30 K), the *P(E)* loops are nearly linear with minimal hysteresis and small polarization values (±3 µC/cm²). The close overlap of successive loops in this temperature range indicates a conventional insulating state without remnant polarization, consistent with the high resistivity (~10$^8$ Ω-m at 30 K, Fig. 12(a)) and spin glass freezing observed below 30 K.

In the intermediate range (30–70 K), the resistivity decreases by several orders of magnitude [Fig. 12(a)], and the *P(E)* loops begin to show slight hysteresis [Fig. 18(b)]. This behavior correlates with the onset of short-range magnetic correlations detected in our ac susceptibility study and literature reported µSR measurements [41]. Raman spectroscopy also points to enhanced spin-lattice coupling in this regime. These signatures suggest that short-range magnetic order promotes the formation of polar nano-regions via spin-lattice interactions. As a result, the system acquires a hybrid dielectric-resistive character, reflecting the frustrated triangular spin-chain geometry of CCIO.

Above 70 K, the *P(E)* response changes markedly. The loops broaden into rounded, resistive-like shapes with anomalously large polarization values (±25 µC/cm²), which deviate from classical ferroelectric switching [Fig. 18(c)]. The concomitant sharp drop in resistance indicates that leakage currents dominate the transport response in this high-temperature regime.

Overall, the evolution of the dielectric response mirrors the magnetic state of CCIO. The low-temperature insulating phase (≤30 K) is governed by spin-glass freezing; the intermediate regime (30–70 K) by short-range cluster-glass like magnetic correlations and spin-lattice coupling; and the high-temperature phase by leakage-dominated conduction. This strong correlation highlights robust magnetodielectric coupling, where magnetic disorder and spin–lattice interactions directly govern the dielectric and transport properties.

The structurally analogous spin-chain oxides, such as $Ca_3Co_2O_6$ [46,93,94], $Ca_3Co_{1.4}Rh_{0.6}O_6$ [19], are known to exhibit field-tunable dielectric properties arising from frustrated magnetic configurations. Pioneering studies have shown that in these materials, frustration-driven magnetic states strongly couple to the lattice, giving rise to complex magnetodielectric phenomena under applied magnetic fields. In CCIO, $P(E)$ measurements conducted in zero applied magnetic field reveal distinct dielectric regimes closely linked to the evolution of magnetic states with temperature. These findings reinforce the central role of magnetic frustration and spin-lattice coupling in stabilizing emergent dielectric states in one-dimensional spin-chain systems, and positions CCIO as a promising platform for exploring magnetoelectric crossovers in geometrically frustrated oxides.

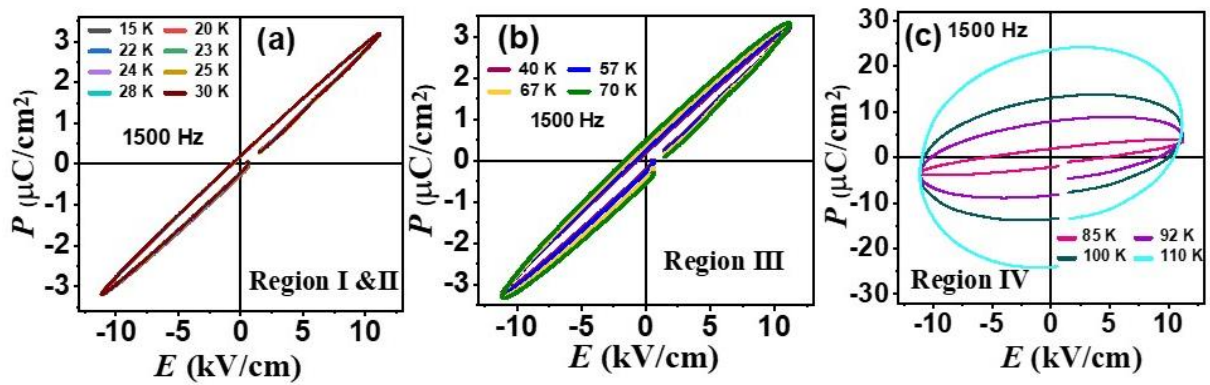

**Figure 18:** Polarization–electric field ($P$–$E$) loops of CCIO measured at ±10 kV/cm and 1500 Hz across different temperature ranges: (a) 15–30 K, showing nearly linear dielectric behavior; (b) 40–70 K, where slight hysteresis develops due to short-range magnetic correlations; and (c) 85–110 K, exhibiting broadened resistive-like loops with large apparent polarization values dominated by leakage currents.

The complete picture of magnetic phase diagram of CCIO is shown in Fig. 19, which reveals a rich sequence of field-tunable states with temperature. At low temperatures (below ~20 K), the system enters a frozen-PDA regime, with metastable, glassy dynamics driven by spin locking and a field-tunable vertical shift in the $M(H)$ loop. A narrow crossover between 20–30 K connects the frozen-PDA state to the higher-temperature PDA regime accompanied by a field-independent horizontal exchange bias. Over 30 K to 100 K, a partially disordered antiferromagnetic (PDA) phase is stabilized; in this regime, high magnetic fields (≥50 kOe) align the otherwise disordered one-third chain sublattice. Above 100 K, the system shows a transition to a short-range correlated regime which persist up to room temperature. This hierarchy of phases reflects the combined effects of geometric frustration, strong anisotropy and spin–orbit coupling, and quasi-one-dimensional lattice topology.

The phase boundaries are cross-validated by transport, thermodynamic, and spectroscopic correlations. The ac susceptibility exhibits a frequency-dependent peak between 100 and 30 K, tracking the gradual onset of glassy, frustrated dynamics within the PDA regime, while the dc anomaly centered near 30 K marks the transition into a frozen cluster-glass state. Under high fields (≥50 kOe), field-aligned reorientation of the disordered one-third chains produces a broad anomaly in $M(T)$ between 55 and 90 K and, in $M(H)$, a field-dependent loop opening despite zero coercivity and remanence at $H \to 0$ for 100 K $\gtrsim T \gtrsim$ 35 K. Specific heat study reveals the same spin alignment as a broad anomaly under magnetic field in the 50–65 K window, and also reveals additional low-energy magnetic excitations below $\approx$ 10 K (Schottky-like contribution), consistent with a frozen PDA/cluster-glass manifold.. The high-resistivity insulating state correlates with the slowing down of spin dynamics, indicating a strong coupling between the spin and charge degrees of freedom. Raman spectroscopy shows pronounced, non-monotonic shifts and linewidth anomalies of $E_g$ modes at the magnetic crossovers, while $A_{1g}$ modes remain structural fingerprints; these phonon anomalies occur without a structural transition confirmed by neutron diffraction, demonstrating that magnetic interactions dominate phonon behavior. Enhanced spectral weight above 1000 cm$^{-1}$ links increased spin disorder and SOC effects to the quantum ground state. Polarization–electric-field loops evolve across the PDA and frozen regimes, establishing intrinsic magnetodielectric coupling that follows the magnetic phase evolution. Together, these mutually consistent signatures define a coherent, field-responsive phase diagram in which PDA order emerges, reorients under field, and freezes into a glassy state, with transport, heat capacity, Raman, and dielectric observables all tracking the same underlying spin-lattice-orbital interplay.

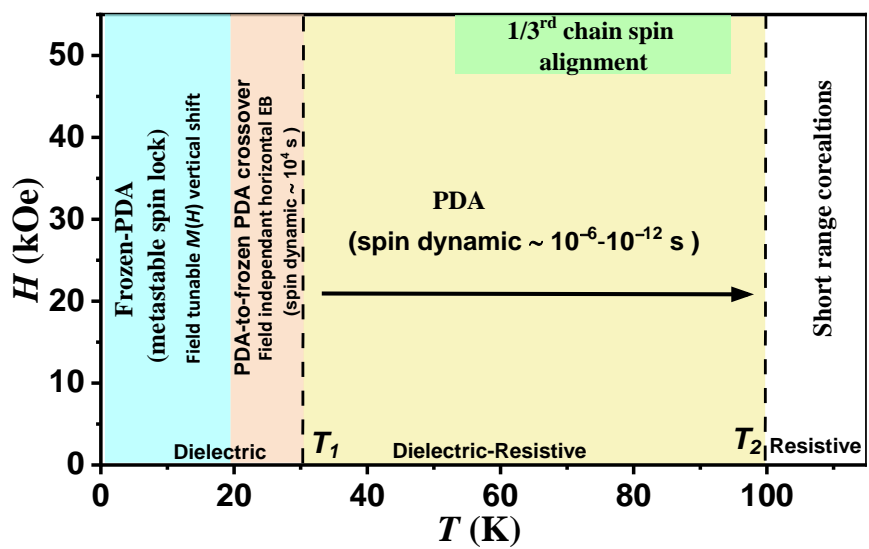

**Figure 19:** Schematic of magnetic phase diagram of CCIO. PDA is the partially disordered antiferromagnetic state and EB is the exchange bias.

## IV.    Summary and Conclusions

We have carried out comprehensive investigation of CCIO using structural, magnetic, thermodynamic, transport, and dielectric probes, including x-ray diffraction, temperature-dependent neutron powder diffraction (NPD), dc magnetization, ac susceptibility, specific heat capacity, resistivity, Raman scattering, and polarization - electric field ($P$ - $E$) measurements. These complementary studies establish the cooperative influence of geometric frustration, spin–orbit coupling, and quasi-one-dimensional lattice topology on the emergent properties of CCIO.

The results show that CCIO undergoes a gradual, field-tunable transformation from a short range correlated state to a partially disordered antiferromagnetic (PDA) state, characterized by persistent glassy spin dynamics, robust exchange bias, and magnetodielectric coupling. DC magnetization, ac susceptibility, and magnetization relaxation measurements reveal a highly glassy magnetic landscape, characterized by slow, cluster-like relaxation and strong aging effects. Under high applied fields (≥30 kOe) a field-aligned reorientation of the disordered one-third chain sublattice sets in below 100 K, manifested as a broad anomaly in $M(T)$, a field-dependent hysteresis opening in $M(H)$ despite zero coercivity and remanence at $H \to 0$ for 100 K $\gtrsim T \gtrsim$ 35 K, and a distinct ac-susceptibility peak between 100 K and 30 K that tracks the evolution of these partially disordered chains. Isothermal $M(H)$ curves show partial spin-chain freezing, while exchange bias (EB) measurements confirm volume-based EB rooted in the PDA structure.

Transport and specific heat studies further evidence strong spin-charge-lattice coupling: resistivity anomalies correlate with spin dynamics, and specific heat reveals both a significant temperature dependent linear term and magnetic anomalies linked to spin-chain dynamics. Raman spectroscopy and $P$-$E$ loops uncover pronounced spin-phonon and magnetodielectric coupling, establishing that magnetic disorder directly governs lattice vibrations and dielectric response throughout the phase diagram.

In summary, CCIO emerges as a model frustrated spin-chain system where geometric frustration, strong spin–orbit coupling, and the quasi-one-dimensional lattice produce a rich spectrum of correlated phenomena, field-tunable partial magnetic order, glassy spin dynamics, exchange bias, and magnetodielectric coupling. These findings sharpen our understanding of partial disorder and frustration in low-dimensional magnets and highlight CCIO as a promising platform for exploring multifunctional and quantum phases under varied external stimuli.

**ACKNOWLEDGMENTS**

PM acknowledges V. B. Jayakrishnan for his support with the X-ray diffraction measurements. PM thanks Kuldeep Singh Chikara for valuable scientific discussions and insightful comments. SMY acknowledges the financial assistance from Anusandhan National Research Foundation (ANRF), Department of Science and Technology, Government of India, under the J C Bose fellowship program (JCB/2023/000014).